\documentclass[letterpaper,titlepage,12pt]{article}

\pdfoutput=1

\usepackage{amssymb,amsmath,amsfonts}
\usepackage{graphicx}
\usepackage{epsfig}
\usepackage{ccaption}

\def\clock{{\count0=\time
           \divide\count0 60
           \ifnum\count0<10 0\fi\the\count0
           \multiply\count0 -60 \advance\count0 \time
           :\ifnum\count0<10 0\fi \the\count0
         }}
\newcommand{\timestamp}{{\small\vbox{\hbox{\tt\jobname.tex}
\hbox{\the\day/\the\month/\the\year, \clock}}}}

\newcommand{\ie}{{\it i.e.,\,}}
\newcommand{\eg}{{\it e.g.,\,}}

\newcommand{\lp}{\left(}
\newcommand{\rp}{\right)}

\newcommand{\beq}{\begin{equation}}
\newcommand{\eeq}{\end{equation}}
\newcommand{\bea}{\begin{eqnarray}}
\newcommand{\eea}{\end{eqnarray}}
\newcommand{\beqa}{\begin{eqnarray}}
\newcommand{\eeqa}{\end{eqnarray}}

\newcommand{\Hf}{$\mathcal{H}_f$}

\numberwithin{equation}{section}

\begin{document}

\begin{titlepage}
\begin{flushright}
\end{flushright}

\vskip 2cm
\begin{center}
{\bf\LARGE{Black String Flow
}}
\vskip 1.2cm
{\bf 
Roberto Emparan$^{a,b}$,
Marina Mart{\'\i}nez$^{b}$
}
\vskip 0.5cm
\medskip
\textit{$^{a}$Instituci\'o Catalana de Recerca i Estudis
Avan\c cats (ICREA)}\\
\textit{Passeig Llu\'{\i}s Companys 23, E-08010 Barcelona, Spain}\\
\smallskip
\textit{$^{b}$Departament de F{\'\i}sica Fonamental and}\\
\textit{Institut de
Ci\`encies del Cosmos, Universitat de
Barcelona, }\\
\textit{Mart\'{\i} i Franqu\`es 1, E-08028 Barcelona, Spain}\\

\vskip .2 in

\end{center}

\vskip 0.3in

\baselineskip 16pt
\date{}

\begin{center} {\bf Abstract} \end{center} 

We give an exact description of the steady flow of a black string into a planar horizon. The event horizon is out of equilibrium and provides a simple, exact instance of a `flowing black funnel' in any dimension $D\geq 5$. It is also an approximation to a smooth intersection between a black string and a black hole, in the limit in which the black hole is much larger than the black string thickness. The construction extends easily to more general flows, in particular charged flows.

\vskip 0.2cm

\noindent 
\end{titlepage} \vfill\eject

\setcounter{equation}{0}

\pagestyle{empty}
\small
\normalsize
\pagestyle{plain}
\setcounter{page}{1}

\newpage

\section{Introduction}

Recent studies of black holes and black branes have sparked an interest in stationary spacetimes admitting event horizons that are not Killing horizons, \ie\ the null generators of the horizon are not parallel to the generators of isometric time translations.\footnote{To the best of our knowledge, horizons with this property were first described, independently, in \cite{Hubeny:2009ru} and \cite{Khlebnikov:2010yt}.} From a technical viewpoint, the theorems \cite{Hawking:1971vc,Hollands:2006rj,Moncrief:2008mr} that would forbid this possibility are evaded since the horizons are non-compactly generated. From a physical perspective, such horizons connect two asymptotic regions of infinite extent which have different surface gravities, \ie\ different temperatures.  They can be regarded as describing a steady heat flow between two infinite heat reservoirs that keep a temperature gradient constant in time.

In this article we describe a remarkably simple, exact solution for a `flowing horizon'. The explicit nature of the construction allows a detailed study of the properties of the flow. Since the spacetime is Ricci-flat, it shows that, in contrast to previous descriptions of horizon flows (motivated by AdS/CFT) \cite{Hubeny:2009ru,Khlebnikov:2010yt,Khlebnikov:2011ka,Fischetti:2012ps,Figueras:2012rb,Fischetti:2012vt}, a negative cosmological constant is not essential for their existence.

In order to motivate the construction, let us first imagine a thin black string, of thickness $r_{bs}$, that falls vertically into a very large black hole of radius $r_{bh}\gg r_{bs}$.\footnote{We may envisage two black strings, falling at antipodal points of the black hole horizon, in order to avoid any total pull on the black hole. Note that the notion of the string falling `vertically' along its length is sensible since the worldsheet of a black string is not Lorentz invariant.} The black string has a much larger surface gravity than the black hole. If the string is free-falling into the black hole horizon, then there are no external forces acting on the system and we can expect that the two horizons merge smoothly.\footnote{Instead of meeting at a singular cusp \cite{Emparan:2011ve}, which in the present context would be unphysical.} This is not a stationary configuration: the black hole is accreting mass from the string that flows into it and therefore must grow in size. However, this effect becomes negligible if we take the limit $r_{bh}\to\infty$ keeping $r_{bs}$ fixed, and focus on the region where the two horizons meet. The black hole horizon then becomes an acceleration, Rindler-type infinite horizon, into which the black string flows by falling  freely across it. Going to the rest frame of the falling black string, the acceleration horizon disappears: we are left with the spacetime of a static black string.

In other words, if we take a static black string and view it from a frame that accelerates along the direction of the string, what we observe is a string in free fall into an acceleration horizon. We will construct the event horizon for such accelerated observers (taking also into account their dragging by the string, as we will see), and show that it interpolates smoothly between the Rindler horizon of a `cold', infinitely-large black hole, at large distances from the string, and the Killing horizon of the `hot' black string when far from the acceleration horizon. It is clear that the spacetime has a timelike Killing vector --- which defines the rest frame of the string --- but the flowing event horizon is not mapped into itself under its action.

The horizon of this `black string flow' is closely similar to the `black funnels' of \cite{Hubeny:2009kz,Hubeny:2009rc,Caldarelli:2011wa,Fischetti:2012ps,Santos:2012he,Figueras:2012rb,Fischetti:2012vt}, where a string-like horizon that in one direction extends towards the AdS boundary, in the other direction smoothly merges with the infinitely extended horizon of an AdS black brane. The two constructions differ in their asymptotics but otherwise describe essentially similar phenomena. Our construction should approximate well the horizon of an AdS funnel much thinner than the AdS radius, in the region where it joins the AdS black brane. Indeed, it should give the universal description of all neutral, non-rotating, thin black funnels over distances sufficiently close to the horizon.

\section{Horizon of black string flow}

In the rest frame of the free-falling black string, and in $D=n+4$ spacetime dimensions, the metric is
\beq\label{bsmetric}
ds^2=-f(r)dt^2 +dz^2 +\frac{dr^2}{f(r)}+r^2 d\Omega_{n+1}\,,
\eeq
with
\beq
f(r)=1-\lp\frac{r_0}{r}\rp^n\,.
\eeq
In the absence of the string ($f=1$), the null surfaces $t= z+t_0$, with constant $t_0$, are (future) acceleration horizons, \ie\ event horizons for trajectories of asymptotically uniform acceleration along $z$. Often this is made manifest by changing to coordinates adapted to accelerating observers, but this is actually not needed, nor is it very practical in the present case. Instead, it is simpler to trace back an appropriate congruence of null rays from null asymptotic infinity.\footnote{An approach similar in essence was previously used in \cite{Figueras:2009iu}.} In our case, one condition that we clearly want to be satisfied is that the null rays reach the conventional Rindler horizon far from the string, \ie\
\beq\label{horcond}
\frac{\dot t}{\dot z}\to 1\quad \mathrm{and}\quad \dot r\to 0\quad \mathrm{for}\quad r\to\infty\,,
\eeq
where the dot denotes derivative with respect to an affine parameter $\lambda$.
These turn out to be the main conditions that we will need to impose, with all other initial conditions on the geodesics following naturally.

The angles on $S^{n+1}$ remain fixed for each null geodesic so that the event horizon preserves the symmetry $SO(n+2)$. The equations for $t(\lambda)$, $z(\lambda)$, $r(\lambda)$  are easy to obtain from
\beq\label{nulleq1}
-f \dot t^2+\dot z^2+\frac{\dot r^2}{f}=0\,,
\eeq
and
\beq\label{nulleq2}
\dot t=\frac{\epsilon}{f}\,,\qquad \dot z=p\,,
\eeq
where $\epsilon$ and $p$ are two integration constants coming from the isometries generated by $\partial_t$ and $\partial_z$. Then \eqref{horcond} is satisfied by setting
\beq
\frac{\epsilon}{p}=1\,.
\eeq

\paragraph{Null hypersurface \Hf.} The null hypersurface ruled by outgoing geodesics can now be characterized by the one-form equation
\beq
dt=dz+\frac{\sqrt{1-f}}{f}dr\,,
\eeq
\ie\
\beq\label{evh}
t=z+t_0+\int \frac{\sqrt{1-f}}{f}dr\,,
\eeq
each value of $t_0$ giving a different null hypersurface that ends at a different value of the null coordinate in future null infinity. Obviously, any of them can serve as our event horizon, differing simply by a translation in $t-z$. We shall denote the null hypersurface with $t_0=0$ as \Hf\ --- the flowing, or funnel, horizon. Clearly it is not a stationary horizon: the action of $\partial_t$ changes $t_0$ and therefore it does not map a hypersurface onto itself but rather onto another one. 
The explicit form of the integral in $r$ in \eqref{evh} is not particularly illuminating and we give it in the appendix. The surface \Hf\ is plotted in fig.~\ref{fig:evh}.

\begin{figure}[t]
\begin{center}
\includegraphics[scale=.9]{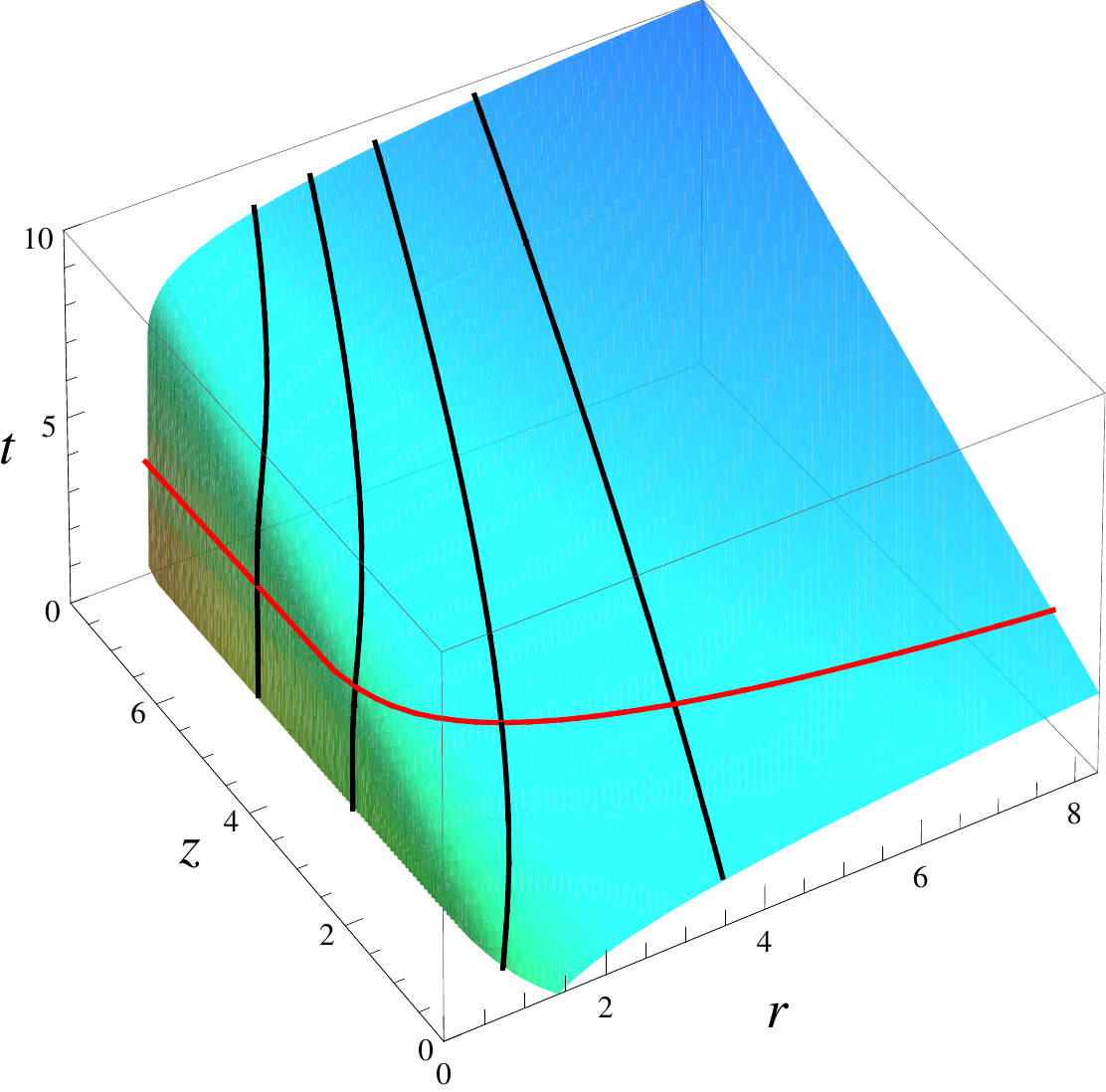}
\end{center}
\caption{\small Event horizon \Hf\ of the black string flow as the hypersurface \eqref{evh} in $(t,r,z)$ space (for $n=2$ and $r_0=1$). It extends along $-\infty<z,t<\infty$ and $1<r<\infty$. The black string lies at $r=1$, and is reached at $t\to-\infty$.  At any finite $r$, the surface tends to $t=z$ at large $z$. The black curves are null geodesics representative of the congruence that rules the hypersurface, for (left to right) $\zeta=5,3,0,-6$. The red curve is a constant-$t$ section.
}\label{fig:evh}
\end{figure}

We can introduce coordinates $x_\pm$ adapted to these null surfaces, defined by
\beq\label{xpm}
dx_\pm=dt\pm \left(dz+\frac{\sqrt{1-f}}{f}dr\right)\,.
\eeq
These are null one-forms normal to the hypersurfaces defined by $dx_\pm=0$. The vectors $\partial/\partial x_\pm=(\partial_t\pm \partial_z)/2$ are instead spacelike vectors tangent to these hypersurfaces, \ie\ $dx_-\cdot (\partial/\partial x_+)=0$ and $dx_+\cdot (\partial/\partial x_-)=0$. 
Moreover, the hypersurfaces $dx_+=0$ (resp.~$dx_-=0$) are symmetric under the action of $\partial/\partial x_-$ (resp.~$\partial/\partial x_+$).

\begin{figure}[t]
\begin{center}
\includegraphics[scale=.9]{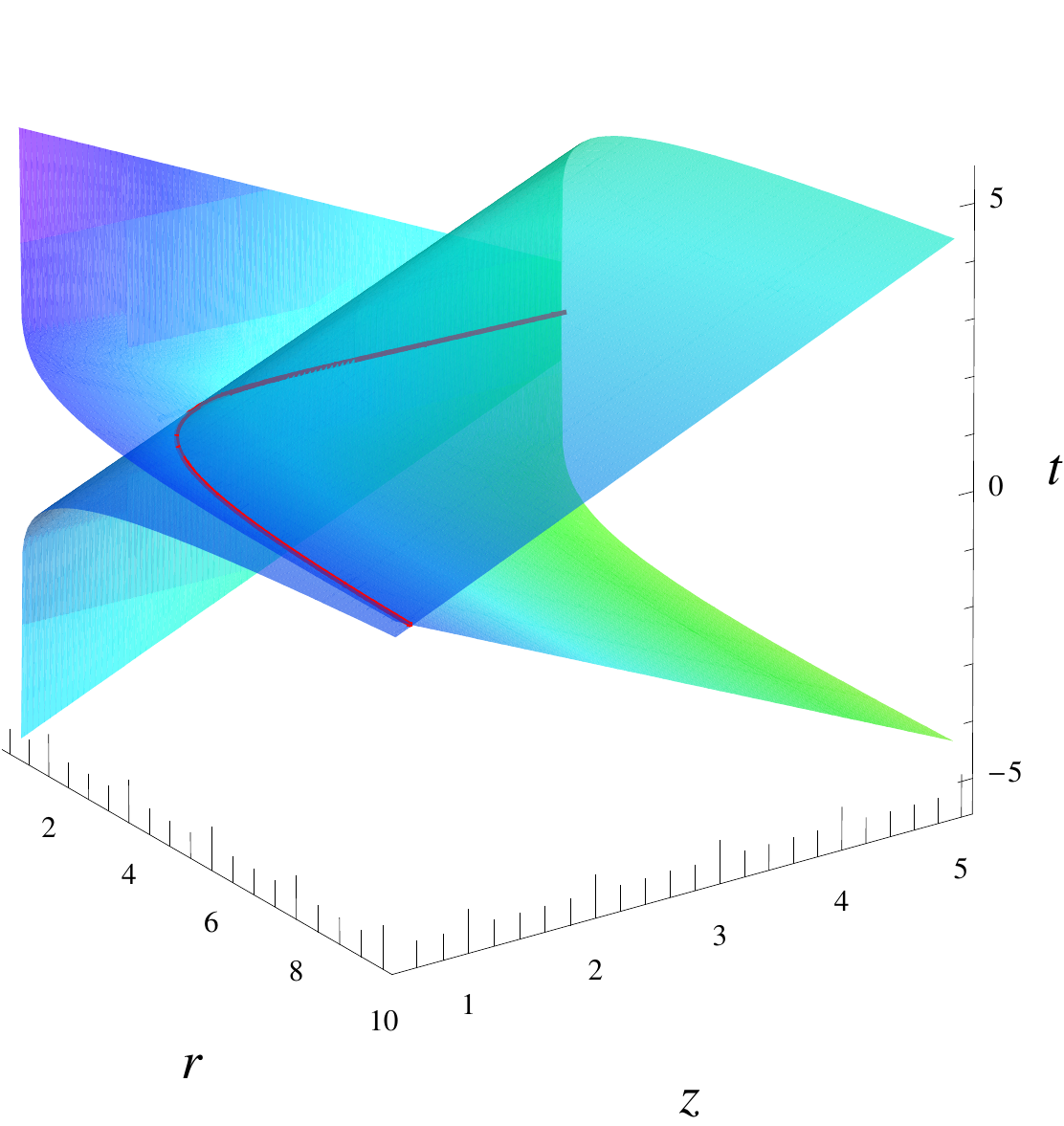}
\end{center}
\caption{\small The wedge formed by the null hypersurfaces $x_+=0$, $x_-=0$ (for $n=3$). The curve at their intersection at $t=0$ is marked in red.
}\label{fig:wedge}
\end{figure}

The geometry \eqref{bsmetric} written
in coordinates $(x_-,z,r,\Omega)$ takes the form
\beq\label{adapmet}
ds^2=-f dx_-^2-2 dx_-\left( f dz+\sqrt{1-f}dr\right)+\lp dr -\sqrt{1-f}\,dz\rp^2
 +r^2 d\Omega_{n+1}\,.
\eeq
\Hf\ is the null surface $x_-=0$. Taking also its time reversal, namely the null surface $x_+=0$, we can regard the region $x_-<0$, $x_+>0$ that they bound as the Rindler wedge modified by the presence of the black string, see fig.~\ref{fig:wedge}.

\paragraph{Null geodesic congruence.} The null geodesics that rule \Hf\ are easily obtained. Using the freedom to rescale $\lambda$ we set $\epsilon=p=1$. Then, since
\beq
\dot z=1\,,\qquad \dot r=\sqrt{1-f}\,,
\eeq
we have
\beqa\label{geozr}
z(\lambda)&=&\lambda+\zeta\,,\notag\\
r^\frac{n+2}{2}(\lambda)&=& r_0^{n/2}\lp r_0+\frac{n+2}{2}\lambda\rp\,,
\eeqa
and $t(\lambda)$ is obtained from \eqref{evhexp}. Here $\zeta$ is an integration constant that labels each null ray of the congruence. It corresponds to the value of $z$ for the ray when $\lambda=0$, \ie\ when $r=r_0$ and $t\to-\infty$. We can then take $(\lambda,\zeta)$ as the coordinates on the null hypersurface (together with the angles of $S^{n+1}$). If we eliminate them we obtain the hypersurface $t(r,z)$ in \eqref{evh}.

For each value of $\zeta$ we have a light ray outgoing in the $r$ direction, which initially hovers just above the black string horizon, until it escapes out to infinity moving in the $r$ and $z$ directions, see fig.~\ref{fig:evh}. 

Note that $r=r_0$ is reached at a finite value of the affine parameter, namely $\lambda=0$.  This is in fact the same as for null outgoing trajectories outside the horizon in the Schwarzschild geometry: they have $\epsilon>0$ and reach $r=r_0$ in the \textit{past} horizon at a finite affine parameter. In our construction the same happens for the null geodesics in \Hf. Taking $\lambda<0$ they are extended into the interior of the white hole until they reach at $r=0$ the past curvature singularity of the solution.

\paragraph{Event horizon and black string drag.} \Hf\ given by \eqref{evh} is a codimension-1 null hypersurface that extends to asymptotic infinity. It is the future null boundary of a region of spacetime, and it is natural to ask what are the timelike trajectories that have \Hf\ as their event horizon.

According to eq.~\eqref{geozr}, all light rays on \Hf\ move towards $r\to\infty$ as the affine parameter grows. Then, any timelike trajectory that remains within bounded values of $r$ will cross \Hf\ at a finite time. That is, \Hf\ is \textit{not} an event horizon for observers that remain within a finite range of the black string: they all fall across \Hf\ eventually. 

We interpret this phenomenon physically as a dragging effect. A boosted black string (one that moves at constant velocity) has around itself an ergosurface at a constant radial distance. Observers inside this surface cannot remain static but are dragged along with the string. In our configuration the black string is accelerating, \ie\ its velocity grows, and so the ergoregion grows too. We expect then that any trajectory that remains within a finite distance from the string will be dragged along with it and eventually cross the acceleration horizon, thus moving to the future of \Hf.
An observer who wants to avoid crossing \Hf\ must not only accelerate in the $z$ direction, but it must also move out towards $r\to\infty$. These are the observers that have \Hf\ as their event horizon.\footnote{The dragging effect becomes weaker in higher dimensions. For very large $n$ it is only appreciable within a distance $r_0/n$ of the black string \cite{Emparan:2013moa}. As $n\to\infty$, the horizon becomes exactly a planar Rindler horizon outside this region.} It is straightforward to extend this analysis to the class of observers whose motion is confined inside the wedge in fig.~\ref{fig:wedge} by considering trajectories which are time-reversal invariant around $t=0$.

\begin{figure}[t]
\begin{center}
\includegraphics[scale=1]{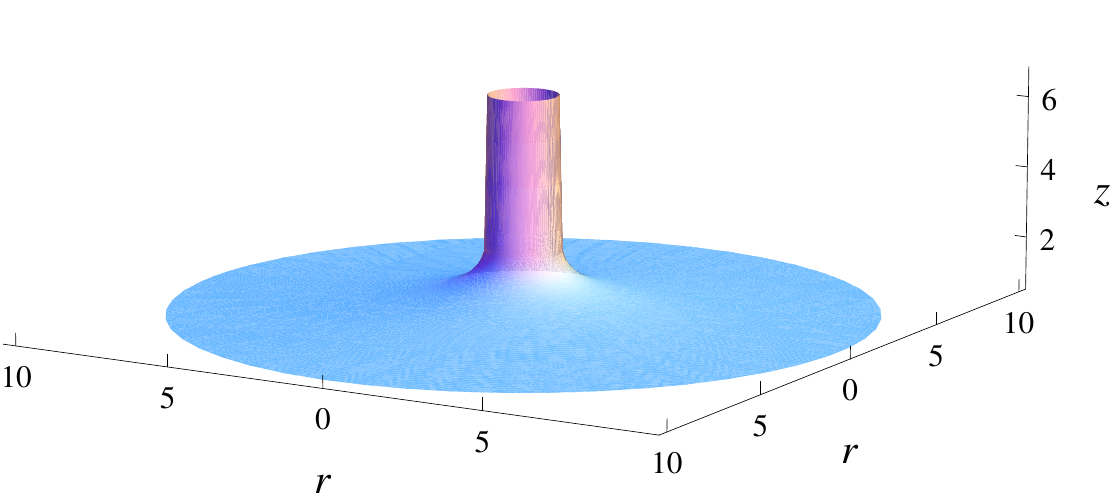}
\end{center}
\caption{\small Constant-time cross section of the event horizon \eqref{crosseh} (for $n=3$ and $r_0=1$), illustrating the funnel-shape that interpolates between the black string and the infinite planar acceleration horizon. The circles at constant $z$ are actually $S^{n+1}$. The funnel extends infinitely in $z$ and $r$.
}\label{fig:funnel}
\end{figure}

\paragraph{Funnel geometry.} Consider a cross-section of this horizon at constant $t$ (or equivalently, at constant $x_+$). From \eqref{xpm} and \eqref{adapmet}, the metric induced on it is
\beq\label{sechor}
ds^2_\mathrm{(hor)}=\frac{dr^2}{f^2}+r^2 d\Omega_{n+1}\,.
\eeq
This geometry describes an infinite funnel: at $r\to\infty$ it becomes flat space, while near $r=r_0$, where $f$ vanishes linearly in $r$, we find an infinite throat with the geometry $\mathbb{R}\times S^{n+1}$, with sphere radius $r_0$.\footnote{Coincidentally, this is the same geometry as the spatial section of the extremal Reissner-Nordstrom solution.}
We can describe this surface as the curve
\beq\label{crosseh}
z+\int \frac{\sqrt{1-f}}{f}dr=0\,,
\eeq
which we represent in  fig.~\ref{fig:funnel}. This illustrates clearly the idea that the black string  and the Rindler horizon merge smoothly into a funnel-shaped horizon.

\section{Out-of-equilibrium flow}

The vector
\beqa
\frac{d}{d\lambda}&=&\dot t \frac{\partial}{\partial t}+\dot z \frac{\partial}{\partial z}+\dot r\frac{\partial}{\partial r}\notag\\
&=& \frac1{f}\frac{\partial}{\partial t}+ \frac{\partial}{\partial z}+\sqrt{1-f}\frac{\partial}{\partial r}
\,.
\eeqa
is an affine generator of the null geodesic congruence. It is convenient to consider the following non-affine generator of the future event horizon, 
\beqa
\ell&=&\frac12 f\lp r(\lambda)\rp \frac{d}{d\lambda}\notag\\
&=&\frac12\lp\frac{\partial}{\partial t}+f \frac{\partial}{\partial z}+f\sqrt{1-f}\frac{\partial}{\partial r}\rp\,.
\eeqa
This is normalized in such a way that near and far from the black string we recover the generators of the black string horizon and of the acceleration horizon. Often the Rindler horizon generator is taken to be the boost vector $z\partial_t+t\partial_z$, which at $t=z$ becomes $t(\partial_t+\partial_z)$. However, this is not adequate for us: the boost vector gives a finite, dimensionless surface gravity $\kappa=1$, to the Rindler horizon. This is the acceleration of observers at unit proper distance from the horizon, and not the surface gravity in the infinite radius limit of a black hole, which is zero.

The surface gravity $\kappa_{(\ell)}$ of $\ell$ is defined as the non-affinity factor of the geodesics,
\beq
\nabla_\ell\ell=\kappa_{(\ell)}\, \ell\,.
\eeq

Since $\lambda$ is an affine parameter we easily find that
\beqa
\kappa_{(\ell)}&=&\ell^\mu\partial_\mu \ln f\notag\\
&=& \frac{n}{2r}\lp\frac{r_0}{r}\rp^{3n/2}\,.
\eeqa
This surface gravity decreases monotonically from its asymptotic value at the black string horizon at $r=r_0$, where $\kappa_{(\ell)}\to n/(2r_0)$, down to $\kappa_{(\ell)}\to 0$ at large $r$ where the horizon approximates the planar Rindler horizon. 

Since the horizon is out of equilibrium, we can expect that its expansion be positive. In order to compute it, consider the geometry of sections at constant $\lambda$,
\beq
ds^2_\mathrm{(hor)}=\bigl( 1-f(r(\lambda))\bigr) d\zeta^2+r^2(\lambda) d\Omega_{n+1}\,.
\eeq
Here we use the coordinate $\zeta$ on the surface, instead of $r$ as in \eqref{sechor}, since $\partial/\partial r$ does not commute with $d/d\lambda$ and therefore is not a good coordinate for the congruence. The area element on this surface is
\beq
a=\sqrt{1-f}\,r^{n+1}\omega_{n+1}
\eeq
(where $\omega_{n+1}$ is the area element of $S^{n+1}$)
and therefore the expansion of $d/d\lambda$ is
\beq
\theta_{(\lambda)}=\frac{d\ln a}{d\lambda}=\frac{n+2}{2}\frac{\sqrt{1-f}}{r}\,.
\eeq
This is indeed positive, so the area of the horizon grows to the future. It is also monotonically decreasing, vanishing at $r\to\infty$. 

If we consider the expansion associated to $\ell$ we get
\beq
\theta_{(\ell)}=\ell^\mu\partial_\mu\ln a = \frac12 f\theta_{(\lambda)}=\frac{n+2}{4}f\frac{\sqrt{1-f}}{r}\,.
\eeq
This is again positive, but now it vanishes both at $r=r_0$ and at $r\to\infty$. Thus the flow of the vector $\ell$ reflects the property that this event horizon interpolates between two asymptotic horizons, each of which is asymptotically in equilibrium at a different temperature.

Finally, note that not only is the horizon out of global equilibrium, \ie\ has non-constant $\kappa$, but it is also away from \textit{local} equilibrium. By this we mean that the gradient $r_0\,\partial_z \ln\kappa_{(\ell)}$ becomes large in the region $r\gg r_0$. Then the surface gravity at a section of the horizon at constant $z$, with radius $r(\lambda)$, is not well approximated by the surface gravity of a black string with that horizon radius --- as should be clear from fig.~\ref{fig:funnel}. As a consequence, the flowing horizon cannot be described in the effective hydrodynamic theory for black strings \cite{Emparan:2009at,Camps:2010br}.

\section{Charged flows}

Our previous analysis can be easily extended to more general static black string metrics of the form
\beq
ds^2=-T(r)dt^2+Z(r) dz^2+\frac{dr^2}{R(r)}+r^2 H(r) d\Omega_{n+1}\,,
\eeq
where all the metric functions are assumed positive outside the black string horizon and tend to $1$ at $r\to\infty$. By a suitable choice of the radial coordinate we could set $H=1$, or instead $R=T$. Each choice has its virtues, so we shall keep this radial gauge freedom. 

Solutions with $T=Z$ are qualitatively different from those with $T<Z$ (and when $T>Z$ there are no null geodesics with $\epsilon/p=1$). When $T=Z$ the string worldsheet is Lorentz-invariant and the notion of the string falling along its length is not well defined. The black string horizon does not merge with the Rindler horizon, but instead the two just intersect. This can be easily seen by performing the conventional change to Rindler coordinates, $t=\rho\sinh\eta$, $z=\rho\cosh\eta$: the black string horizon at $r=r_0$, where $T(r_0)=0$, and the Rindler horizon at $\rho=0$, form two intersecting null surfaces, both with zero expansion. Later we discuss a relevant instance of this.

Henceforth we restrict ourselves to $T<Z$. The flowing event horizon is characterized by
\beq
dt=dz+ dr\,\sqrt{\frac{Z(r)-T(r)}{T(r)Z(r)R(r)}}\,,
\eeq
and in terms of the coordinates $(\lambda,\zeta)$ on the congruence, where $\zeta$ labels different null rays and $\lambda$ the affine parameter along them, we have
\beq
dz=Z^{-1}d\lambda+d\zeta\,,\qquad
dr=\sqrt{\frac{R(Z-T)}{TZ}}\,d\lambda\,.
\eeq
The metric on a constant-$t$ section (or constant $\lambda$) of this horizon is
\beqa
ds^2_\mathrm{(hor)}&=&\frac{Z}{TR}dr^2+r^2 H(r) d\Omega_{n+1}\nonumber\\
&=&(Z-T)d\zeta^2+r^2 H(r) d\Omega_{n+1}\,.
\eeqa

For the non-affine null geodesic generator
\beq
\ell=\frac12\lp\partial_t+\frac{T}{Z}\partial_z+\sqrt{(Z-T)\frac{RT}{Z}}\;\partial_r\rp
\eeq
the surface gravity is
\beq
\kappa_{(\ell)}=\sqrt{(Z-T)\frac{R}{TZ}}\;\frac{\partial_r T}{2}\,.
\eeq
This vanishes as $r\to\infty$, while close to $T=0$ it reproduces the surface gravity of the event horizon of the black string, 
\beq\label{bskappa}
\kappa_{(\ell)}\to\sqrt{\frac{R}{T}}\;\frac{\partial_r T}2\,.
\eeq

The expansion is
\beq
\theta_{(\ell)}=\frac12 \lp\frac{\partial_r Z-\partial_r T}{Z-T}+\frac{n+1}{r}+\frac{n+1}2\frac{\partial_r H}{H}\rp \sqrt{(Z-T)\frac{RT}{Z}}\,.
\eeq
The first term inside the brackets is due to the expansion along the string direction, while the latter two correspond to the spherical expansion in the radial direction. The last factor comes from $\ell^\mu\partial_\mu r$.

A natural class of solutions to study are charged strings, in particular electrically charged ones. The qualitative properties differ depending on whether the charge is string-charge, \ie\ the strings are electric sources of a 2-form potential $B_{\mu\nu}$, or 0-brane charge, which sources a Maxwell 1-form potential $A_\mu$. 

\paragraph{String charge.}
Configurations with string charge are of interest for several reasons. The neutral black string flow of previous sections is unstable, since the spacetime \eqref{bsmetric} is itself unstable \cite{Gregory:1993vy}. However, string-charged black strings that are sufficiently close to extremality, but not necessarily extremal, are stable.

An interesting instance are black strings with fundamental string charge, \ie\ black F-strings. Above extremality the horizon can be regarded as the gravitational description of a thermal spectrum of excitations on a stack of fundamental strings. The `F-string flow' horizon then describes, in gravitational terms, the flow of these excitations down a very large black hole that the string intersects.
Even if the F-string charge allows to tune down the temperature of the black string, only at extremality can it be in thermal equilibrium with the infinitely large black hole. This extremal limit has Lorentz-invariance along $z$, with $T=Z$, so in this case there is actually no flow. Above extremality the string excitations are at a higher temperature than the black hole, and the system appears to differ from those in which the string excitations are in thermal equilibrium with a finite-temperature horizon (as studied in a worldsheet approach, \eg\ \cite{Lawrence:1993sg,Frolov:2000kx,Kaya:2002ue,deBoer:2008gu}, or in the blackfold approach \cite{Grignani:2010xm,Grignani:2011mr}). In our construction, when the black string is not extremal it is not mining energy from the black hole, but rather dumping it.

\paragraph{0-brane charge.}
0-brane charge on a black string breaks Lorentz symmetry on the worldsheet at any temperature, including at extremality. The horizons of these strings can then merge smoothly with the Rindler horizon and there is always a non-zero flow. For extremal string flows the surface gravity associated to $\ell$ is zero only at the black string, where \eqref{bskappa} vanishes, and at the Rindler horizon at $r\to\infty$. Inbetween them, the surface gravity is generically non-zero, as is also the expansion $\theta_{(\ell)}$. So these are always out-of-equilibrium configurations.

One may wonder what drives the flow when both its endpoints are at zero temperature. It is easy to see that it is driven by a gradient of the electric potential, \ie\ an electric field along the horizon. The charge on the string is in free fall across the acceleration horizon. On the event horizon, this phenomenon is a charge current from the black string to the planar horizon, driven by an electric field. This field on the event horizon is the projection (pullback) of the spacetime electric field that the static black string creates. Clearly this field points in the direction of increasing $r$, and thus, on the event horizon, it points from the black string towards the planar horizon. It may be interesting to understand better these charge flows. In particular the appearance of a temperature on the horizon of the extremal string flow, in which the asymptotic endpoints are at zero temperature, is suggestive of resistive (Joule) dissipation of the electric current on the horizon.

Finally, 0-brane charge does not prevent the instability of the black string, as this charge can be redistributed along the horizon. However, the addition of string charge can make these solutions stable, even supersymmetric. In the latter case, the flowing horizon is not parallel to the timelike vector associated to the Killing spinors, and therefore need not be an extremal horizon.

\section{Outlook}

The black string flow studied above approximates a system where a very thin black string smoothly pierces a black hole. A similar-looking configuration has been found in the late-time evolution of the black string instability \cite{Lehner:2010pn} --- including a flow from the string that makes the black hole grow. It would be interesting to study in more detail the geometry in the latter case, near the region where the `black hole' and `black string' meet, to see if it conforms to the flowing horizon we have constructed.

Black funnels in AdS can be interpreted holographically in dual terms as a flow of Hawking radiation in the boundary theory, emitted from a black hole through a thermal radiation fluid that extends to infinity \cite{Hubeny:2009ru,Fischetti:2012ps}. For our flowing geometries, a similar interpretation is also possible --- although only to some extent, since the quantum degrees of freedom of the dual radiation are not known. In order to understand how this works, consider first the Rindler horizon, without the string. If we impose Dirichlet boundary conditions on a timelike surface $\mathcal{S}$ at a fixed, finite proper distance from the horizon, then the gravitational dynamics of the system can be described in terms of a dual thermal `Rindler fluid' on $\mathcal{S}$ \cite{Bredberg:2010ky}. If we introduce the black string, then there will be a black hole horizon on $\mathcal{S}$ where it intersects the black string. The dual description, in terms of the quasilocal stress-energy tensor on $\mathcal{S}$, will then exhibit a flow of the Rindler fluid qualitatively similar to that in AdS. Note also that in this Rindler-fluid set up, the C-metric yields an exact droplet solution in a four-dimensional bulk. The construction is like that of a black hole on a thin, planar domain wall in \cite{Emparan:2000fn}. 

The method we have employed of finding non-equilibrium acceleration horizons in stationary black hole spacetimes can be extended to other situations of interest, for instance: (i) rotating black strings, to yield rotating string flows; (ii) black strings in AdS, to obtain black funnels in (hyperbolic) AdS black branes; (iii) Schwarzschild black holes, to find the event horizon in the final plunge of extreme-mass-ratio black hole collisions. We plan to report on these systems elsewhere.

\section*{Acknowledgments}

We gratefully acknowledge conversations with Pau Figueras, Veronika Hubeny, and Don Marolf. Work supported by MEC FPA2010-20807-C02-02, AGAUR 2009-SGR-168 and CPAN CSD2007-00042 Consolider-Ingenio 2010.

\appendix

\section{Explicit integration of the event horizon}
\label{}

Setting for simplicity $r_0=1$, the event horizon \eqref{evh} is the surface
\beqa\label{evhexp}
t-t_0&=&z-\int\frac{dr}{r^{n/2}-r^{-n/2}}\notag\\
&=&z-\frac{2r^{(2-n)/2}}{n-2}\, {}_2 F_1\lp \frac{n-2}{2n},1,\frac{3n-2}{2n};r^{-n}\rp\,.
\eeqa
When $n=2$ we find
\beq
t-t_0=z+\frac12\ln(r^2-1)\,.
\eeq
The expression simplifies for other values of $n$, \eg\
\beq
t-t_0=z+2\sqrt{r}+\ln\frac{\sqrt{r}-1}{\sqrt{r}+1}\qquad (n=1)\,,
\eeq
\beq
t-t_0=z+\frac12 \arctan r+\frac14 \ln\frac{r-1}{r+1}\qquad (n=4)\,.
\eeq
At large values of $r$ the surface tends to
\beq
t-t_0\to z-\frac{2r^{(2-n)/2}}{n-2}\,,
\eeq
so for larger $n$ the horizon asymptotes more rapidly to $t=z$. In fact for $n=1,2$, the limit of $r\to\infty$ at fixed $z$ or fixed $t$ does not tend to $t=z$ (although it is always the case that $dt\to dz$ at  $r\to\infty$). The interpretation is that, as might be expected, low-codimension black string flows spread much more in the transverse directions than higher-codimension flows. Nevertheless, the spatial geometry of the horizon \eqref{sechor} is asymptotically flat in all dimensions.


\end{document}